RESEARCH ARTICLE

# An entrainment-based model for annular wakes, with applications to airborne wind energy

Sam Kaufman-Martin  |  Nicholas Naclerio  |  Pedro May  |  Paolo Luzzatto-Fegiz

Department of Mechanical Engineering, University of California, Santa Barbara, Santa Barbara, CA, USA

**Correspondence**
Sam Kaufman-Martin, Department of Mechanical Engineering, University of California, Santa Barbara, Santa Barbara, CA, USA.
Email: s_kaufman-martin@ucsb.edu

**Abstract**

Several novel wind energy systems produce wakes with annular cross-sections, which are qualitatively different from the wakes with circular cross-sections commonly generated by conventional horizontal-axis wind turbines and by compact obstacles. Since wind farms use arrays of hundreds of turbines, good analytical wake models are essential for efficient wind farm planning. Several models already exist for circular wakes; however, none have yet been proposed for annular wakes, making it impossible to estimate their array performance. We use the entrainment hypothesis to develop a reduced-order model for the shape and flow velocity of an annular wake from a generic annular obstacle. Our model consists of a set of three ordinary differential equations, which we solve numerically. In addition, by assuming that the annular wake does not drift radially, we further reduce the problem to a model comprising only two differential equations, which we solve analytically. Both of our models are in good agreement with previously published large eddy simulation results.

## 1 | INTRODUCTION

Wind turbine power is determined primarily by the diameter of the turbine and its efficiency. This has led to the development of large horizontal axis wind turbines (HAWTs), with diameters of over 130 m.[1] While HAWTs have been very successful in land-based and shallow-offshore arrays, they have yet to meaningfully take advantage of far-offshore wind energy resources.[2] Additionally, HAWTs cannot access wind resources in the upper atmosphere, where the wind power density is about two times greater than it is at typical HAWT hub heights because wind speed increases with altitude.[3]

To take advantage of these currently untapped energy resources, new airborne wind energy (AWE) technologies are being developed that can harvest wind power from a tethered kite or aircraft.[4,5] Several AWE designs consist of a large, tethered kite that flies transverse to the prevailing wind in a closed loop, allowing it to reach the faster-moving winds present higher in the atmosphere. The kite harvests energy either by a turbine generator on the kite that transmits electrical power to the base, or by transmitting mechanical power from the kite to the base. Many different designs have been developed with both of these power mechanisms, as well as with moving or stationary base stations and a variety of flight patterns.[5,6] A commonly considered operating mode involves flying the kite along a circular path, thereby generating a wake with an annular cross-section (henceforth referred to as an annular wake); this is the case, for example, for the Makani energy kite.[7] As AWE development continues, the technology may eventually be viable for large-scale wind farms. Indeed, conceptual studies have estimated that power generated by AWE systems could potentially satisfy a significant portion of global power demand.[8] However, for large-scale AWE farms to become a reality, an understanding of their wakes will be critical for determining array spacing and layout.

Surprisingly, while circular wakes have been studied extensively in both the literature on free shear flows[9–11] and in the literature on wind turbine wakes,[12-14] there appear to be no theoretical models of turbulent annular wakes, to the best of our knowledge. The only existing study on annular turbine wakes appears to be a large eddy simulation of AWE kite aerodynamics by Haas and Meyers.[15] Other related works include a study by the same group on pumping-mode AWE devices,[16] a non-turbulent wake model of pumping-mode AWE kites,[17] a CFD study of HAWT wakes with radially-varying thrust distributions,[18] studies on the behavior of annular jets,[19,20] investigations on the wakes behind toroidal bluff bodies at low Reynolds number,[21] and a study on the impact of kites on HAWT farm wakes.[22] Additionally, we note that models of annular wakes, which are the subject of this paper, are distinct from models of circular wakes composed of annular elements, such as the vortex ring model of Øye.[23]

In this paper, we develop two theoretical models for annular wakes by leveraging the concept of entrainment velocity, whose history includes applications to modeling plumes,[24] wakes with circular cross-sections,[25] gravity currents,[26] and wind farms.[27,28] In Section 2, we derive two entrainment-based models for the shape and speed of an annular wake as a function of distance behind an annular obstacle. The first model must be solved numerically, whereas the second can be solved analytically. In Section 3, we then compare the models to the simulation results from Haas and Meyers,[15] finding good agreement. A discussion is presented in Section 4, with conclusions following in Section 5.

## 2 | ENTRAINMENT-BASED MODELS OF ANNULAR WAKES

### 2.1 | Assumptions and definitions for models

For simplicity, it is assumed that the obstacle that generates the wake is perpendicular to a constant oncoming wind with a steady, uniform velocity $V_\infty$, and that the shape of this wake is statistically axisymmetric and time-independent (the validity of these assumptions will be addressed at the end of this section). The distance between the inner and outer radii of the obstacle is denoted by $S$, and the overall diameter of the obstacle is $D$, as shown in Figure 1A,B. The obstacle is assumed to slow the airflow immediately downstream of it, according to the predictions of actuator disc theory. Unlike in a conventional (circular) axisymmetric wake, there will be a core region at the center of the annulus-shaped wake whose axial velocity $V_i$ will initially be greater than the wake velocity $V_w$. The initial value of $V_i$ should be close to $V_\infty$ (although not necessarily identical) and may vary with distance $x$ behind the turbine. It is assumed that the wake can be completely described by the wake velocity $V_w$, wake span $S_w$, and total diameter $D_w$, which depend only on $x$. Turbulent entrainment is expected to increase $V_w$, $S_w$, and $D_w$ with increasing $x$.

In order to model the effects of turbulent entrainment, we employ the well-established "entrainment hypothesis." This turbulence closure was first introduced by G.I. Taylor in the context of modeling turbulent plumes.[24] As noted earlier, this turbulence closure has been applied extensively in geophysical and industrial flow problems including plumes,[29] jets,[25] natural ventilation,[30] and gravity currents,[31] as well as wind turbines and wind farms.[27,28]

Following the entrainment hypothesis, fluid downstream of the obstacle is assumed to entrain into the wake with a radial velocity that is proportional to the streamwise velocity difference across the interface. This yields entrainment from the external flow with a radial velocity of $w_e = E(V_\infty - V_w)$ and from the core region with a radial velocity of $w_i = E(V_i - V_w)$, where $E$ denotes the entrainment coefficient (see Figure 1B,C). The relationship between $E$ and Reynolds stresses, as well as other physical quantities, has been described in previous studies.[28,31,32] In brief, $E$ constitutes a Reynolds stress that has been made nondimensional using appropriate mean velocity scales.

Similarly to other established wake models,[33] the pressure in the wake is assumed to be constant (after the initial adjustment that takes place immediately behind the obstacle) and equal to the ambient pressure (see Section 2.5). The assumption that the wake grows by turbulent entrainment is valid as long as the flow is turbulent and isobaric, regardless of the shape of the wake.

Although the assumption that the average wake flow is time-independent and axisymmetric is well-established in low-order models of HAWT wakes,[12,13,27] it is worth discussing whether or not these assumptions are still valid when modeling the wake of an AWE device. First, we note that one-dimensional momentum theory (which underpins actuator disc theory) makes no assumptions about the rotor design of the wind-energy-extracting device,[34] which means it is valid even for one-bladed HAWTs. Since the AWE device modeled in this paper is analogous to a detached turbine blade, this supports using the actuator disc assumption here. Second, we note that the outer wingtips of AWE kites can have similar tip speed ratios ($\lambda$) to conventional HAWTs. For example, the AWE device modeled by[15] has a tip speed ratio of $\lambda = 7$, which is also a common value for HAWTs.[34] This information can be used to estimate the pitch of the helical wake generated by the device. For a wind-energy-harvesting device sweeping a circular path perpendicular to the oncoming wind, the period $\mathcal{T}$ of the turbine's rotation can be defined as

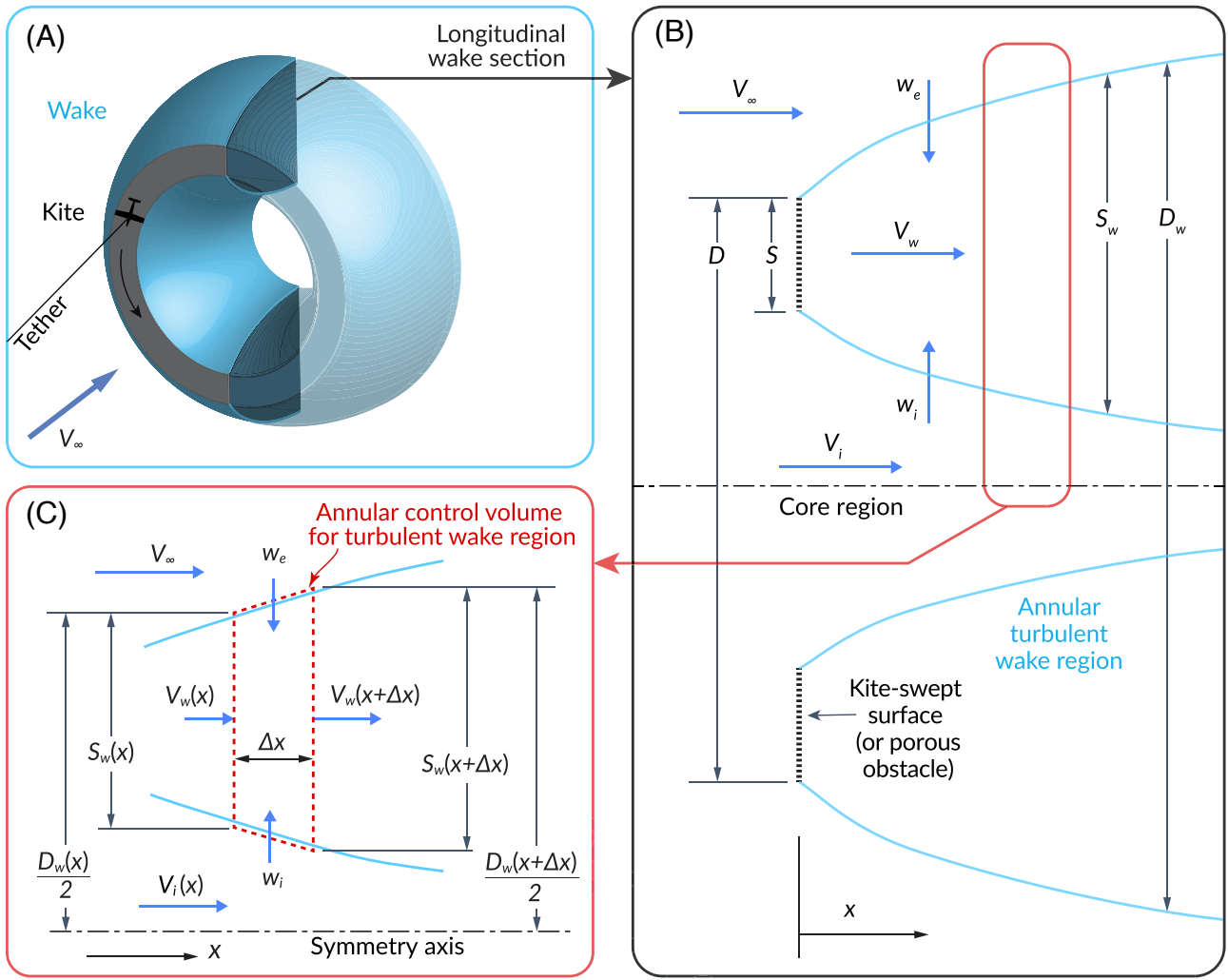

**FIGURE 1** (A) Schematic of the development of an annular wake behind a porous obstacle or kite flying on a circular path. (B) Annular-wake cross-section, showing definitions for the entrainment model. (C) Cross-section of the annular control volume used for the turbulent wake region; for clarity, only the top portion is shown. A similar control volume analysis is used for the core region [Colour figure can be viewed at wileyonlinelibrary.com]

$$\mathcal{T} = \frac{\pi D}{\lambda V_\infty}. \quad (1)$$

Using this definition, we can describe the pitch $h$ of the helical wake formed by such a device in terms of known quantities:

$$h = V_\infty \mathcal{T} = \frac{\pi D}{\lambda}. \quad (2)$$

From this equation, we can see that for any device with $\lambda = 7$, the helical wake will have $\frac{h}{D} < 0.5$. Given the tightness of the helix generated by such an AWE device, the time-averaged flow in the wake may be approximated as statistically axisymmetric. This further validates both the use of actuator disc theory to model the near-wake behavior and the assumption of axisymmetric growth of the wake by turbulent entrainment.

## 2.2 | Full model: Flux conservation in annulus and core

Equations for mass and momentum conservation were derived for the annular wake and the core region, following an approach analogous to previous analyses of circular wakes.[25,27] With reference to the wake sketched in Figure 1A, consider an axisymmetric control volume of infinitesimal length $\Delta x$ in the axial direction, with radial dimensions corresponding to the inner and outer radial boundaries of the wake at the given distance $x$, as sketched in Figure 1C. In a given plane, the area of the core region is $\frac{\pi}{4}(D_w - 2S_w)^2$, implying that the area of the annular wake is $\frac{\pi}{4}[D_w^2 - (D_w - 2S_w)^2]$, which can be simplified to $\pi S_w (D_w - S_w)$. Assuming the fluid has uniform density, conservation of mass in the annular control volume implies that

$$\pi S_w (D_w - S_w) V_w + \pi D_w w_e \Delta x + \pi (D_w - 2S_w) w_i \Delta x = \pi S_w (D_w - S_w) V_w + \frac{d}{dx}[\pi S_w (D_w - S_w) V_w] \Delta x. \tag{3}$$

Dividing through by $\pi \Delta x$, canceling the first term on each side, and using the definition of the entrainment velocity given in Section 2.1, the equation for conservation of mass in the annulus is

$$\frac{d}{dx}[S_w (D_w - S_w) V_w] = E(V_\infty - V_w) D_w + E(V_i - V_w)(D_w - 2S_w). \tag{4}$$

Assuming a negligible pressure gradient in the wake, the above approach can then be used to derive an equation for conservation of axial momentum in the annulus. The momentum flow into the control volume in Figure 1C must equal the momentum flow exiting the control volume. To obtain the $x$-direction momentum fluxes, we take the corresponding mass flux terms shown in Equation (3) and multiply them by the relevant axial velocities (which represent $x$-momentum per unit mass). This gives the following equation for momentum conservation in the annular control volume:

$$\pi S_w (D_w - S_w) V_w^2 + \pi V_\infty D_w w_e \Delta x + \pi V_i (D_w - 2S_w) w_i \Delta x = \pi S_w (D_w - S_w) V_w^2 + \frac{d}{dx}\left[\pi S_w (D_w - S_w) V_w^2\right] \Delta x. \tag{5}$$

Once again, we divide through by $\pi \Delta x$, cancel the first term on each side, and substitute in the definition of the entrainment velocity. This gives the following equation for the conservation of momentum in the $x$-direction in the annulus:

$$\frac{d}{dx}\left[S_w (D_w - S_w) V_w^2\right] = E V_\infty (V_\infty - V_w) D_w + E V_i (V_i - V_w)(D_w - 2S_w), \tag{6}$$

Using the same approach employed to obtain (4) and (6), it is then possible to derive the equations for conservation of mass and momentum in the core region (Equations 9 and 10). With $\pi$ times the density factored out, the conservation equations are as follows, with (4) and (6) repeated first for clarity:

$$\frac{d}{dx}[S_w (D_w - S_w) V_w] = E(V_\infty - V_w) D_w + E(V_i - V_w)(D_w - 2S_w), \tag{7}$$

$$\frac{d}{dx}\left[S_w (D_w - S_w) V_w^2\right] = E V_\infty (V_\infty - V_w) D_w + E V_i (V_i - V_w)(D_w - 2S_w), \tag{8}$$

$$\frac{d}{dx}\left[\frac{1}{4}(D_w - 2S_w)^2 V_i\right] = -E(V_i - V_w)(D_w - 2S_w), \tag{9}$$

$$\frac{d}{dx}\left[\frac{1}{4}(D_w - 2S_w)^2 V_i^2\right] = -E V_i (V_i - V_w)(D_w - 2S_w). \tag{10}$$

Upon closer examination, one can show that only three differential equations and one constant are necessary to fully describe the mass and momentum fluxes in this model. Using the product rule to expand the left side of Equation (10) gives

$$\frac{d}{dx}\left[\frac{1}{4}(D_w - 2S_w)^2 V_i\right] V_i + \frac{1}{4}(D_w - 2S_w)^2 V_i \frac{dV_i}{dx} = -E V_i (V_i - V_w)(D_w - 2S_w). \tag{11}$$

Multiplying Equation (9) by $V_i$ and subtracting it from Equation (11) then gives

$$\frac{1}{4}(D_w - 2S_w)^2 V_i \frac{dV_i}{dx} = 0, \tag{12}$$

which implies

$$V_i = \text{constant}. \tag{13}$$

Therefore, Equations (9) and (10) are equivalent, and we retain only Equations (7), (8), and (9) to solve for the three remaining unknowns $V_w$, $D_w$, $S_w$.

In order to solve this system of differential equations numerically, a set of initial conditions (ICs) is required. Therefore, ICs are derived by applying streamtube analysis and actuator disc theory to the flow in Figure 1. Similar to approaches for circular wakes,[13] here we use one-dimensional momentum theory[34] to derive the initial values of $V_i$, $V_w$, $S_w$, and $D_w$ at $x = 0$. The wake-generating force is thereby accounted for through the initial value of the wake velocity immediately behind the turbine. From the definition of the axial induction factor $a$ in actuator disc theory, it follows that $V_{w,0} = V_\infty(1 - 2a)$. The area of the annulus can then be obtained through mass conservation in the streamtube passing through the obstacle:

$$\frac{S_{w,0}(D_{w,0} - S_{w,0})}{S(D-S)} = \frac{1-a}{1-2a}. \tag{14}$$

Since there is no induction in the core region, one-dimensional momentum theory predicts that $V_{i,0} = V_\infty$. Once again, applying mass conservation in the core region provides the initial condition for the core area:

$$D_{w,0} - 2S_{w,0} = D - 2S. \tag{15}$$

Equations (14) and (15) can be combined to solve explicitly for $D_{w,0}$ and $S_{w,0}$. Thus, the ICs for the system of ordinary differential equations are as follows:

$$V_{w,0} = V_\infty(1 - 2a), \tag{16}$$

$$D_{w,0} = \sqrt{D^2 + S(D-S)\frac{4a}{1-2a}}, \tag{17}$$

$$S_{w,0} = S + \frac{1}{2}(D_{w,0} - D). \tag{18}$$

The fact that $V_i = \text{constant} = V_{i,0} = V_\infty$ allows us to further simplify Equations (7)–(9) as follows:

$$\frac{d}{dx}[S_w(D_w - S_w)V_w] = 2E(V_\infty - V_w)(D_w - S_w), \tag{19}$$

$$\frac{d}{dx}\left[S_w(D_w - S_w)V_w^2\right] = 2EV_\infty(V_\infty - V_w)(D_w - S_w), \tag{20}$$

$$\frac{d}{dx}\left[\frac{1}{4}(D_w - 2S_w)^2 V_\infty\right] = -E(V_\infty - V_w)(D_w - 2S_w). \tag{21}$$

While the initial value problem (IVP) represented by (16)–(21) can be numerically solved as-is with software such as Wolfram Mathematica, one can simplify the equations by restating them in terms of the mass and momentum fluxes in the wake and core, such that each left-hand side in (19)–(21) is the derivative of a single dependent variable, rather than a combination of $S_w$, $D_w$, and $V_w$ (this is a common approach for entrainment-based models in geophysics[25]). In the following definitions, $m$ signifies mass flux and $M$ denotes momentum flux. Once again, $\pi$ times the density has been factored out:

$$m_w \equiv S_w(D_w - S_w)V_w, \tag{22}$$

$$M_w \equiv S_w(D_w - S_w)V_w^2, \quad (23)$$

$$m_i \equiv \frac{1}{4}(D_w - 2S_w)^2 V_\infty. \quad (24)$$

Using the above definitions, one can rewrite Equations (19)–(21) in the following form:

$$\frac{dm_w}{dx} = 2E\left(V_\infty - \frac{M_w}{m_w}\right)\left[\sqrt{\frac{m_i}{V_\infty} + \frac{m_w^2}{M_w}} + \sqrt{\frac{m_i}{V_\infty}}\right], \quad (25)$$

$$\frac{dM_w}{dx} = V_\infty \frac{dm_w}{dx}, \quad (26)$$

$$\frac{dm_i}{dx} = -2E\left(V_\infty - \frac{M_w}{m_w}\right)\sqrt{\frac{m_i}{V_\infty}}. \quad (27)$$

The IVP represented by Equations (16)–(27) was integrated using MATLAB's ode45 solver, which determined the resolution in $x$ automatically to stay within the solver's solution error tolerance of $10^{-8}$. The solution was obtained within a fraction of a second, as is to be expected for a quasi-one-dimensional model. The following definitions (rearranged versions of Equations 22–24) were then used to convert the results back into the desired form:

$$V_w = \frac{M_w}{m_w}, \quad (28)$$

$$S_w = \sqrt{\frac{m_i}{V_\infty} + \frac{m_w^2}{M_w}} - \sqrt{\frac{m_i}{V_\infty}}, \quad (29)$$

$$D_w = 2\sqrt{\frac{m_i}{V_\infty} + \frac{m_w^2}{M_w}}. \quad (30)$$

Thus, results were obtained for $V_w$, $S_w$, and $D_w$, as a function of $x$ (see Section 3).

## 2.3 | Analytical model: No Radial Wake Drift

While the full model produced results that were a close match to the simulations of Haas and Meyers[15] (see Section 3), we were curious to see if an analytical model of the wake could be derived via a slight simplification of the control volume analysis. Therefore, a second model was derived using a different approach to the behavior of the core region, inspired by the qualitative appearance of the wake profile in the results of Haas and Meyers.[15] Instead of using the conservation equation (21) to model the core region, the center of the wake ring (located at radial distance $\frac{D_w}{2} - \frac{S_w}{2}$ from the axis of turbine rotation) is assumed to not drift from its initial position at $\frac{D_{w,0}}{2} - \frac{S_{w,0}}{2}$. Equivalently, this assumes $D_w - S_w = D_{w,0} - S_{w,0} = $ constant, which is henceforth called the no radial wake drift assumption.

There are theoretical and empirical reasons supporting the use of this No Radial Wake Drift approximation in deriving an engineering model for an annular wake. Theoretically, because of the axisymmetry of the problem, the continuity equation would require that there be a significant change in $V_i$ in order for radial displacement of the wake to occur. In our derivation in § 2.2, $V_i$ is found to be constant and equal to $V_\infty$, since the fluid in the core does not pass through the kite-swept surface and does not lose momentum. Empirically, we observed in the results of Haas and Meyers[15] (see Figures 2 and 3 further below) that the annular wake expands at similar rates into the outer flow and into the core region, as long as the diameter of the core region is larger or similar in magnitude to the span of the annular wake, as is expected theoretically.

The governing equations for this model are mass and momentum conservation in the annulus (i.e., Equations (19) and (20) from the full model) and the No Wake Drift assumption:

$$\frac{d}{dx}[S_w(D_w - S_w)V_w] = 2E(V_\infty - V_w)(D_w - S_w), \quad (31)$$

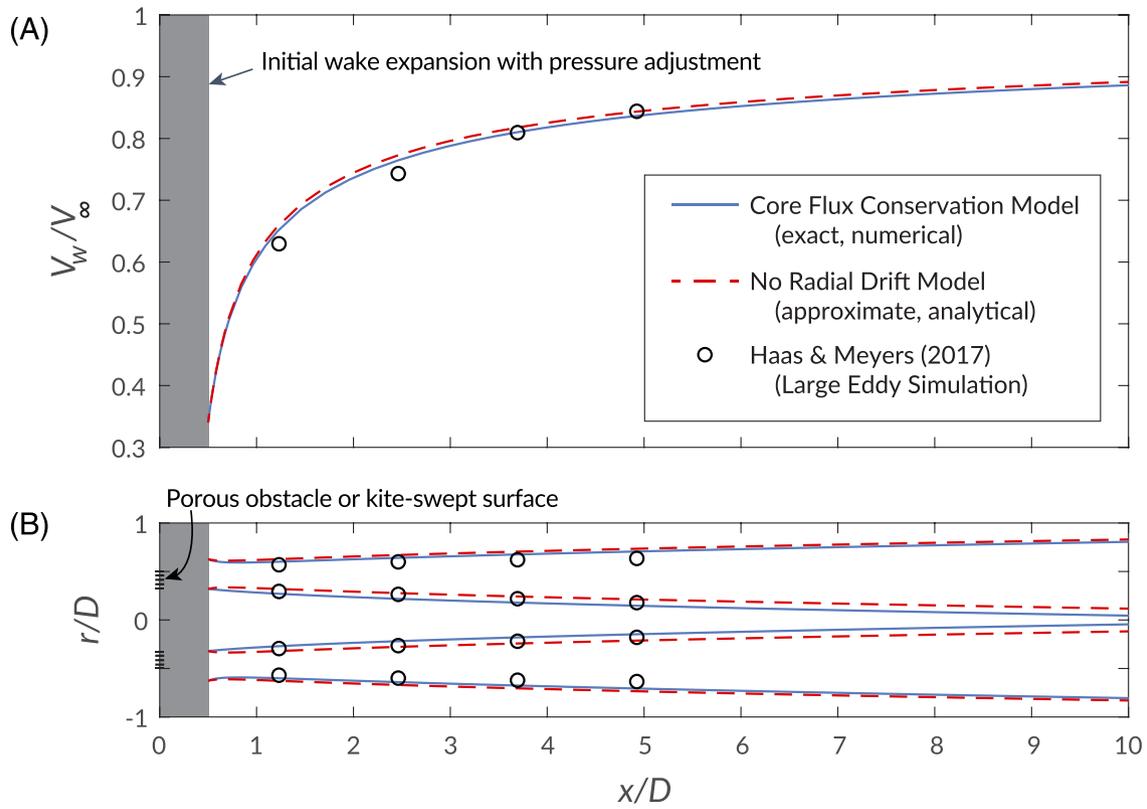

**FIGURE 2** Laminar inflow case, comparing entrainment wake models and the simulation of Haas & Meyers.[15] (A) Wake velocity. (B) Wake boundary. Outer curves are $\pm \frac{D_w}{2}$; inner curves are $\pm(\frac{D_w}{2} - S_w)$ (equivalent to $\pm\sqrt{\frac{m_i}{V_\infty}}$) [Colour figure can be viewed at wileyonlinelibrary.com]

$$\frac{d}{dx}\left[S_w(D_w - S_w)V_w^2\right] = 2EV_\infty(V_\infty - V_w)(D_w - S_w), \quad (32)$$

$$D_w - S_w = D_{w,0} - S_{w,0}. \quad (33)$$

Since we assumed that $D_w - S_w$ is a constant, this term can be factored out of (31) and (32). Multiplying (31) by $V_\infty$ and subtracting (32) gives

$$\frac{d}{dx}[S_w V_w(V_\infty - V_w)] = 0. \quad (34)$$

Integrating, we find that

$$S_w V_w(V_\infty - V_w) = C, \quad (35)$$

where $C$ is a constant. A value for $C$ can be obtained by using the same ICs (16)–(18) already obtained in Section 2.2:

$$C = S_{w,0} V_\infty^2 2a(1 - 2a). \quad (36)$$

Solving Equation (35) for $S_w$ and substituting it into (31), one can integrate the resulting equation, solve for $V_w$, and find that

$$V_w = V_\infty - \sqrt{\frac{C}{4E(x - x_c)}}, \quad (37)$$

where $x_c$ is a constant of integration, which is found to be

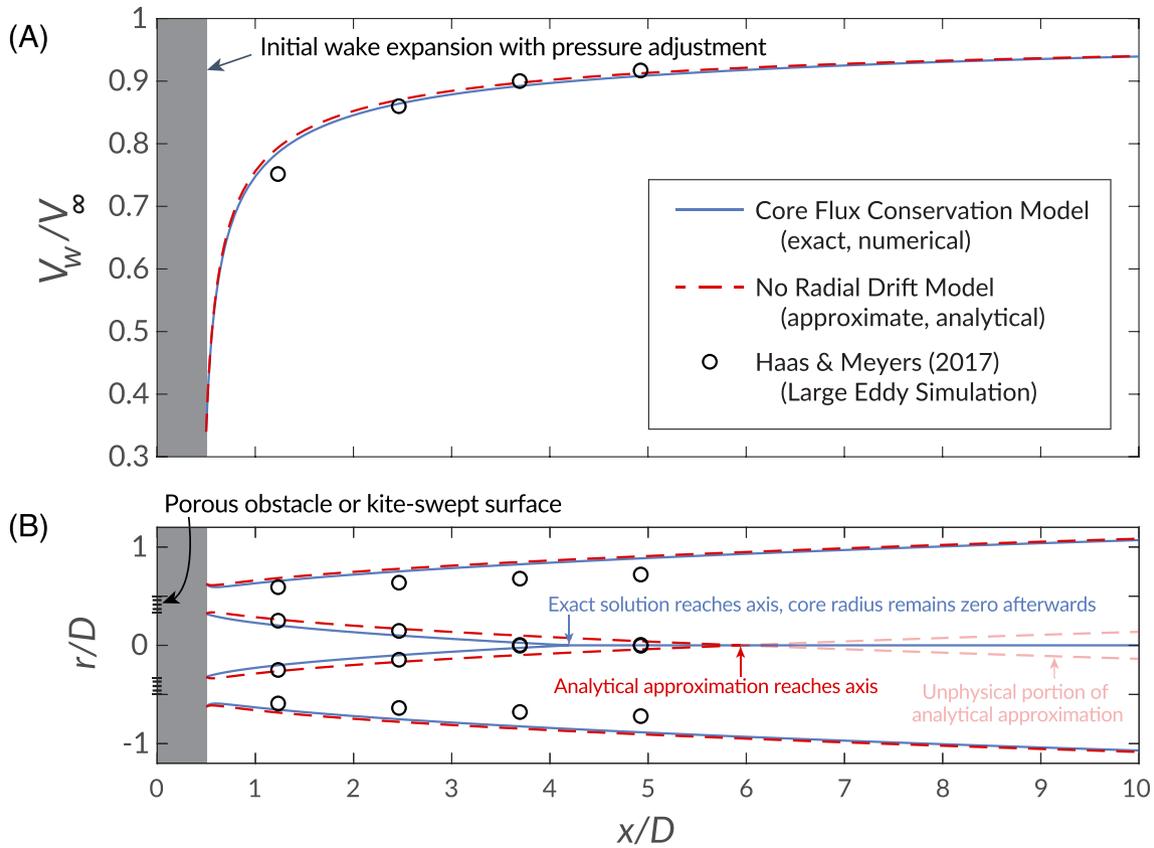

**FIGURE 3** Turbulent inflow case, comparing entrainment wake models and the simulation of Haas and Meyers.[15] (A) Wake velocity. (B) Wake boundary. Outer curves are $\pm \frac{D_w}{2}$; inner curves are $\pm\left(\frac{D_w}{2} - S_w\right)$ (equivalent to $\pm\sqrt{\frac{m_i}{V_\infty}}$) [Colour figure can be viewed at wileyonlinelibrary.com]

$$x_c = -\frac{S_{w,0}(1-2a)}{8Ea}. \tag{38}$$

Given this result for $V_w$, the above equations can be combined to solve for $S_w$ and $D_w$.

In summary, the No Radial Drift assumption gives a set of three equations that form a complete analytical model of an annular wake's evolution. The full set of equations is listed below:

$$V_w = V_\infty \left[1 - \sqrt{\frac{4a^2}{\frac{8Eax}{S_{w,0}(1-2a)}+1}}\right], \tag{39}$$

$$S_w = S_{w,0}\frac{2a(1-2a)V_\infty^2}{V_w(V_\infty - V_w)}, \tag{40}$$

$$D_w = D_{w,0} - S_{w,0} + S_w. \tag{41}$$

## 2.4 | Predicting core disappearance

An additional feature of the analytical model derived in the previous section is that it allows one to immediately predict the downstream location at which the core region of an annular wake disappears (i.e., where $\frac{D_w}{2} - S_w = 0$), which we call $x_{nc}$ ("x no-core"). Downstream of this location, the analytical, entrainment-based model for circular wakes[27] can be used to describe the wake without any further loss of accuracy associated with the No Radial Drift assumption. Furthermore, the expression for $x_{nc}$ can provide insight about the sensitivity of the model to the value of the entrainment coefficient $E$, as described below.

To derive an equation for $x_{nc}$, we start by substituting the condition for core disappearance ($D_w = 2S_w$) into (33) to get an expression for $S_w(x_{nc})$ in terms of the known initial wake geometry:

$$S_w(x_{nc}) = D_{w,0} - S_{w,0}. \tag{42}$$

We then substitute $S_w(x_{nc})$ from (42) into Equation (40) and rearrange the terms to obtain

$$V_w(V_\infty - V_w) = \frac{S_{w,0}}{D_{w,0} - S_{w,0}} 2a(1-2a)V_\infty^2 \text{ for } x = x_{nc}. \tag{43}$$

The right hand side is a constant for a given set of initial conditions; denoting it as $c_1$ for brevity, we rearrange the left-hand side to obtain a quadratic equation for $V_\infty - V_w$:

$$(V_\infty - V_w)^2 - V_\infty(V_\infty - V_w) + c_1 = 0 \text{ for } x = x_{nc}. \tag{44}$$

This quadratic has solutions:

$$V_\infty - V_w = \frac{V_\infty \pm \sqrt{V_\infty^2 - 4c_1}}{2} \text{ for } x = x_{nc}. \tag{45}$$

The physically relevant sign for the "±" is a minus sign; for example, this choice gives $V_w = V_\infty$ when $a = 0$, as expected. We now note that Equation (39), which gave $V_w(x)$, can also be rearranged in terms of $V_\infty - V_w$:

$$V_\infty - V_w = V_\infty \sqrt{\frac{4a^2}{\frac{8Eax}{S_{w,0}(1-2a)} + 1}}. \tag{46}$$

Finally, setting Equations (45) and (46) equal to each other, we solve for $x_{nc}$ to obtain:

$$x_{nc} = \frac{S_{w,0}(1-2a)}{E} \left( \frac{2a}{\left[1 - \sqrt{1 - \frac{S_{w,0}}{D_{w,0} - S_{w,0}} 8a(1-2a)}\right]^2} - \frac{1}{8a} \right). \tag{47}$$

Having obtained an analytical equation for $x_{nc}$, we may also obtain an equation for the velocity in the wake at $x_{nc}$. Substituting $x_{nc}$ from Equation (47) into (39) and simplifying gives

$$V_w(x_{nc}) = \frac{V_\infty}{2} \left[ 1 + \sqrt{1 - \frac{S_{w,0}}{D_{w,0} - S_{w,0}} 8a(1-2a)} \right]. \tag{48}$$

Additionally, the wake diameter at $x_{nc}$ can be obtained by substituting $S_w(x_{nc})$ from Equation (42) into (33) and solving for $D_w$, which gives

$$D_w(x_{nc}) = 2(D_{w,0} - S_{w,0}). \tag{49}$$

It is notable that in this model, the velocity and diameter of the wake at the location of core disappearance are independent of the entrainment coefficient. Indeed, the normalized quantities $\frac{V_w(x_{nc})}{V_\infty}$ and $\frac{D_w(x_{nc})}{D}$ depend only on the initial geometry (i.e., $S/D$) and thrust properties ($a$) of the annular obstacle. Of course, the location of $x_{nc}$ is highly dependent on $E$, as shown by (47).

## 2.5 | Expansion length

As mentioned in Section 2.1, the entrainment hypothesis assumes that pressure in the wake is equal to the ambient pressure. In actuality, wind turbines cause a pressure drop immediately behind the rotor. Actuator disc theory predicts that, over a short distance behind the rotor (henceforth referred to as the expansion length), the wake expands as the fluid continues to decelerate until the pressure in the wake equalizes with the ambient pressure.[34] At this point, the regime of turbulent entrainment is expected to become dominant in the wake.

To account for the expansion length in our models, the results of both calculations are shifted forward in the x direction by an empirically-chosen constant $x_e$ with values typically in the range of $0 < x/D < 1$. This is conceptually similar to the introduction of a virtual origin in entrainment-based plume models.[35,36]

## 3 | RESULTS

### 3.1 | Numerical simulations of Haas and Meyers[15]

We compare the models obtained here with the recent large eddy simulations of Haas and Meyers,[15] who considered both laminar and turbulent inflow conditions. To facilitate this comparison, velocity profiles from figure 5 in Haas and Meyers[15] were converted into tophat velocity profiles in the following manner. (Note that $U_\infty$ in Haas and Meyers[15] is equivalent to $V_\infty$ in this paper.) Data points with $u_x/U_\infty < 1$ are considered to be inside the wake; therefore, the surface where $u_x/U_\infty = 1$ is considered to be the wake boundary. The normalized wake velocity $V_w/V_\infty$ is considered to be the area average of $u_x/U_\infty$ values inside the wake.

Note that Haas and Meyers[15] defined R as the radial distance to the midpoint of the kite, whereas in this paper, the outer radius of the kite's flight path is used. Therefore, D in this paper is equal to $2R + S$ in Haas and Meyers.[15] The data of Haas and Meyers[15] used in Figures 2 and 3 have been scaled to match the coordinate system defined in Figure 1A.

### 3.2 | Parameters for models and results

The key parameters (independent variables) used as inputs for our models were made to be the same as in Haas and Meyers,[15] that is, $S/D = 0.18$ and $a = 0.33$.

Our models also use the empirical parameter E, which is typically around 0.15 for HAWT models in low-turbulence inflow[27] but can reach up to 0.6 for high-turbulence plumes and jets,[35] and the expansion length $x_e$, which is typically in the range of $0 < x/D < 1$. When choosing values of E and $x_e$, only the fit of the models' predictions with the velocity data from Haas and Meyers[15] was considered, as opposed to attempting to fit both the wake shape and the velocity. This is because the "wake boundary" is an artificially-defined parameter compared with the wake velocity, and the wake velocity is the most important parameter for wind farm planning. To model the laminar inflow case from Haas and Meyers,[15] both models use $E = 0.15$ and $x_e = 0.5D$. For the turbulent inflow case, both models use $E = 0.5$ and $x_e = 0.5D$. A comparison between our models and Haas and Meyers[15] is shown in Figures 2 (laminar) and 3 (turbulent).

Additionally, we examine the sensitivity of our models' predictions to the entrainment coefficient E and to the geometry of the kite-swept annulus (S/D) (all using $a = 0.33$). Figure 4A shows how the location at which the core of the wake disappears, $x_{nc}$, depends on E and S/D as predicted by the No Radial Drift (analytical) model (Equation 47). We find that $x_{nc}/D$ decreases as E and S/D increase, as would be expected on physical grounds. Figure 4B demonstrates that, for a given S/D, the approximate value of $x_{nc}$ predicted by the analytical model is directly proportional to the value predicted by the Core Flux Conservation (full) model; this proportionality constant is independent of E. Therefore, the $x_{nc}$ result from the full model can be obtained by multiplying the analytical result by a constant, which depends on S/D but is independent of E. Finally, Figure 4C shows how $V_w(x_{nc})$ varies with S/D in each model.

## 4 | DISCUSSION

In the region of the wake for which the entrainment hypothesis is valid ($x/D \gtrsim 1$), the velocity predictions of both the Core Flux Conservation Model (full model) and the No Radial Drift Model (analytical model) show good agreement with the simulation data from Haas and Meyers[15] in both the laminar and turbulent inflow cases. There appears to be little difference between the velocity results of the two models.

The main difference between the two solutions is that the full model correctly captures the wake-merging behavior in the turbulent case, whereas the analytical model does not (see Figure 3). In the full model, once the core radius ($\frac{D_w}{2} - S_w$) reaches $r = 0$ (i.e., the wake's annulus merges and eliminates the core region), it remains at $r = 0$ for all subsequent values of x/D. In contrast, in the analytical model, the core radius continues shrinking past $r = 0$ and becomes negative at large x/D. This is illustrated by the translucent portion of the innermost dashed lines in Figure 3B. Therefore, the user of the analytical model must know to either disregard the core radius results after the core disappears or, preferably, switch to the circular wake model of Luzzatto-Fegiz[27] (as explained in the next paragraph). However, the position of the outer wake

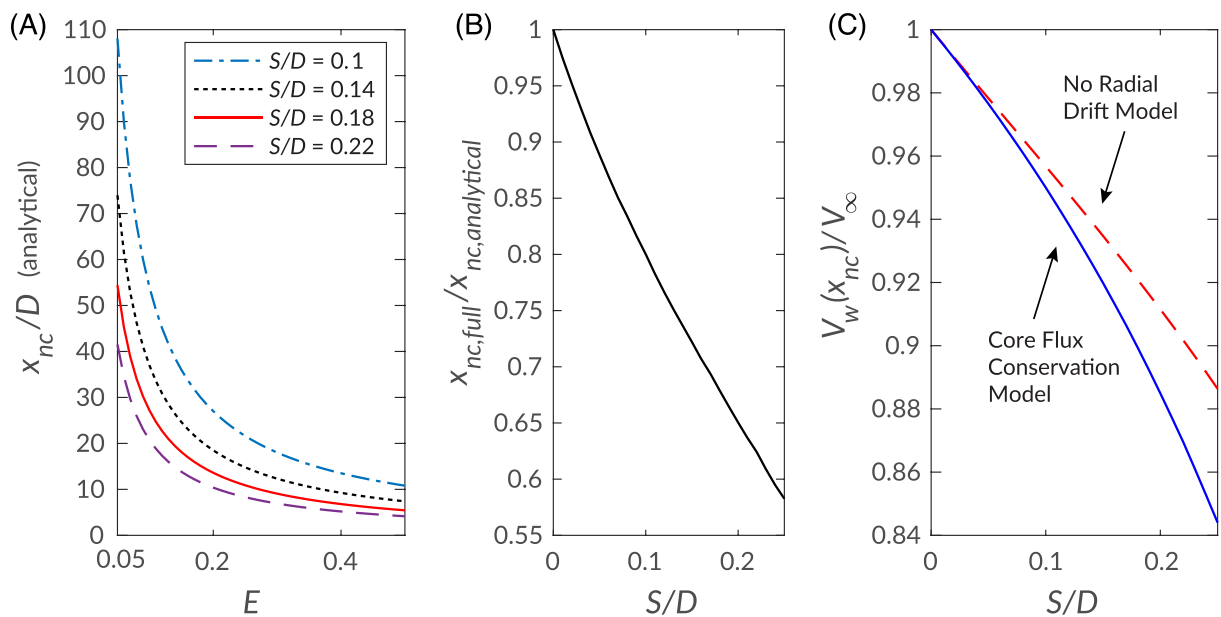

**FIGURE 4** (A) Normalized downstream distance behind an annular obstacle at which the core region disappears ($x_{nc}$), as predicted by Equation (47) from the No Radial Drift Model, shown from $0.05 \leq E \leq 0.5$ and for four different annulus geometries. Note that, for clarity, the initial expansion length $x_e$ is not included in the results as plotted here (that is, $x_e = 0$); in applications, one should add a relevant value of $x_e$. (B) The ratio between the $x_{nc}$ predictions of the Core Flux Conservation (full) and No Radial Drift (analytical) models, plotted versus annulus geometry ($S/D$). (C) Wake velocity ($V_w$) at the location of core disappearance ($x_{nc}$) for both models, plotted versus $S/D$ [Colour figure can be viewed at wileyonlinelibrary.com]

boundary ($\frac{D_w}{2}$) is still reasonably accurate for large $x/D$, as shown by the outermost dashed lines in Figure 3B. This difference between the models is unsurprising given that the Core Flux Conservation Model conserves mass in the core region, whereas the No Radial Drift Model ignores core mass conservation in favor of a geometrically simplified approach that permits an analytical solution.

When the core region is eliminated as the wake annulus merges, one can switch to the traditional, analytical wake model,[25,27] using the values of $V_w$ and $D_w$ at merger as the new initial conditions. For the full annular wake model, this would save computational effort while retaining accuracy, which could be valuable in wind farm optimization applications. Indeed, the results of the full model downstream from the location of core disappearance are equivalent to the analytical model of circular wakes in Morton[25] and Luzzatto-Fegiz.[27]

Although the analytical model is less accurate than the full model, the fact that it does provide a closed-form expression for $x_{nc}$ makes it useful for elucidating some of the properties of annular wakes. For example, in Equation (47) of the analytical model, it is found that $x_{nc}$ is inversely proportional to $E$; solving the numerical model over the full range of expected values of $E$ and $S/D$ demonstrates that this holds true for the full model as well (see Figure 4A,B). Likewise, both models show that the wake velocity at the location of core disappearance ($V_w(x_{nc})$) is independent of $E$ (Figure 4C). That is to say, while the downstream location of $x_{nc}$ depends on the amount of turbulence in the flow, the wake velocity at that location does not. Furthermore, the structures of Equations (19)–(21), (39), and (47) indicate that it is possible to rescale $x$ to a new independent variable $X$, in such a way that $V_w(X)$ and $X_{nc}$ are independent of $E$. Specifically, substituting $X = Ex$ into the above equations makes them independent of $E$. The existence of this "reference solution" $V_w(X)$ suggests that all turbulent annular wakes generated by an annular obstacle with a given $S/D$ are fundamentally similar, and that decreasing the ambient turbulence (parameterized by $E$) serves to simply "stretch" the solution in $x$. Additionally, a reference solution would have benefits for practical wind farm applications, as one would only need to calculate the solution $V_w(X)$ for the full (numerical) model once for a given $S/D$, and then $V_w(x)$ could be calculated for any $E$ simply by transforming $X$ back to $x$. If trying to locate $x_{nc}$, the analytical model could further expedite the process: instead of solving the full numerical model, simply solve Equation (47) and then multiply by the correct constant (which could be tabulated from Figure 4B) to obtain the correct result.

The models introduced here are not restricted to airborne wind energy applications. For example, they could describe the wakes of ground-based annular wind turbines (which have recently drawn attention[37]). With different initial conditions, they could potentially also be applied to the wakes of other toroidal obstacles. The existing initial conditions would be replaced by a matching condition between the momentum deficit and the drag on the obstacle, and a development length that could be found empirically, similarly to approaches for conventional wakes.[33]

## 5 | CONCLUSION

By using the entrainment hypothesis and considering mass and momentum budgets behind an annular obstacle, we derived a model for the spatial development of a turbulent annular wake. Initial conditions were obtained from one-dimensional momentum theory. The model consists of three coupled ordinary differential equations, which are solved numerically. By assuming negligible radial drift in the turbulent wake, we also derived a simplified model, which admits an analytical closed-form solution.

Both models appear to compare well with existing data, especially for larger $x/D$ values. To help provide an even stronger test of the models developed here, experimental or computational results for larger $x/D$ would be helpful, especially given that spacing in wind turbine arrays is typically between $5 \leq x/D \leq 10$.

If supported by additional data from experiments or simulations of annular wakes, the models put forth in this paper would enable the modeling of AWE kite wakes at a much lower computational cost than existing methods. In future work, these models could be combined with existing wake superposition models, such as the momentum-conserving approach developed by Zong and Porté-Agel,[38] in order to represent arrays of multiple kites. This would greatly facilitate the planning of AWE wind farms, which in turn could enable society to harvest previously untapped, energy-dense wind resources.

### ACKNOWLEDGEMENT
We gratefully acknowledge partial support from ARO MURI W911NF-17-1-0306.

### PEER REVIEW
The peer review history for this article is available at https://publons.com/publon/10.1002/we.2679.

### DATA AVAILABILITY STATEMENT
Data sharing is not applicable to this article as no new datasets were generated or analyzed during the current study.

### ORCID
*Sam Kaufman-Martin*  https://orcid.org/0000-0002-2494-161X
*Nicholas Naclerio*  https://orcid.org/0000-0002-7337-3014
*Pedro May*  https://orcid.org/0000-0002-9905-7714
*Paolo Luzzatto-Fegiz*  https://orcid.org/0000-0003-3614-552X